\documentclass[twocolumn,twoside,slac_two]{revtex4}
\usepackage{graphicx}
\usepackage{fancyhdr}
\usepackage{hyperref}
\hypersetup{
    colorlinks=true,
    urlcolor=magenta,
}
\begin{document}

\title{Radio Emission of Air Showers with Energy E$_0$ $\geq$ 10$^{19}$ eV by Yakutsk Array Data}

%

\author{S. Knurenko, I. Petrov, Z. Petrov}
\affiliation{Yu. G. Shafer Institute of Cosmophysical Research and Aeronomy SB RAS, Yakutsk, Russia}

\begin{abstract}
In this paper, we present results obtained from the measurements of radio emission at frequency of 32 MHz with energy more than 10$^{19}$ eV. Generalized formula that describe lateral distribution and depends on main characteristic of the air showers: energy E$_0$ and depth of maximum X$_{max}$ was derived. The formula has a good agreement with data at average and large distances from shower axis. Employing the ratio of radio emission amplitude at distances 175 m and 725 m we determined the depth of maximum X$_{max}$ for air shower with energy 3.7$\cdot$10$^{19}$ eV, which in our case is equal to X$_{max}$ = 769$\pm$34g$\cdot$cm$^{-2}$.
\end{abstract}

\maketitle

\thispagestyle{fancy}

\section {Introduction}

The method of registration of radio emission of ultrahigh-energy particles is based on Askaryan effect. The effect proposes that particle shower develops a negative charge excess, accompanying the passage of ultra-high energy particles through matter [1]. According to this effect, excess of electrons in showers is caused by the annihilation of positrons with electrons of the Earth’s atmosphere. The movement of the shower disc with negative charge excess at a speed greater than the phase velocity of light in that medium is the cause of Cherenkov radiation emission at radio frequencies [2]. In the following years after this discovery, there have been many experimental studies of radio emission from air showers [3, 4]. In paper [7] pointed out the possibility of registration of extensive air showers (EAS) with energies above 10$^{19}$ eV, employing radio equipment placed on the surface of the Earth and registering the radio emission by satellites on the Earth orbit.

In recent years, interest in the air shower radio emission, as an independent method to study the physics of the EAS has grown significantly, and for registration of radio emission were built arrays of significant size [12, 13]. This method makes it possible not only to evaluate the energy, but also to reconstruct the longitudinal shower development, namely, the depth of maximum X$_{max}$ [14, 15]. This is especially important for huge arrays where the uncertainty in the estimation of shower energy with different methods of detecting air showers reaches about (20-40)$\%$. For example, Auger and Telescope Array difference is 20$\%$and the cause of differences is still remains unknown [16]. Thus, the radio emission, in conjunction with other methods of could be employed for intercalibration of huge arrays.

\section{Observation of air showers radio emission of ultrahigh energies}

\begin{figure}
\includegraphics[width=0.8\linewidth]{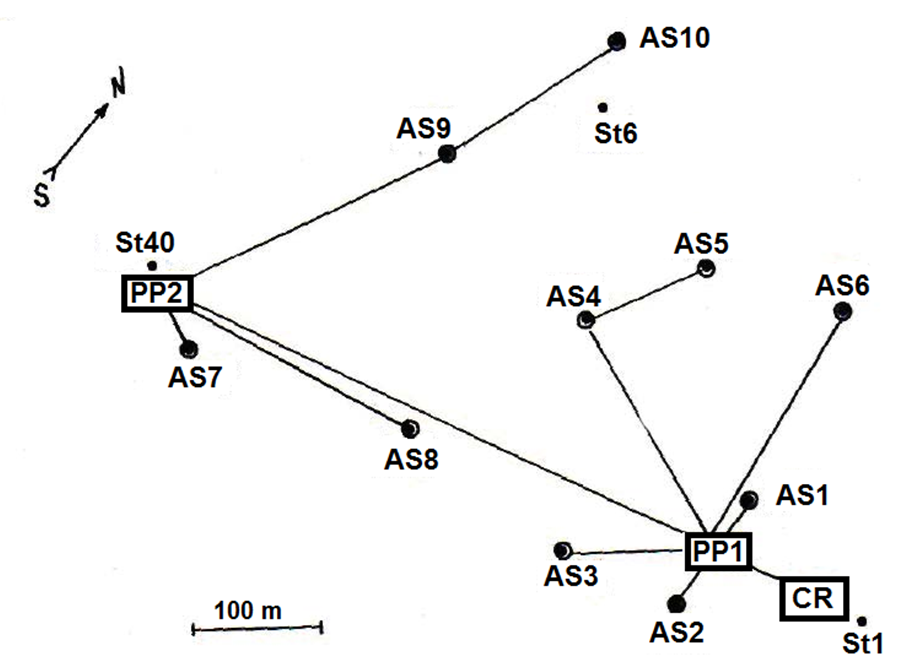}
\caption{Arrangement of the radio antennas, in 1987 - 1989. AS - antenna station; PP - peripheral (intermediate) collection point for air shower data; CR - central registration point; ST – station with scintillation detectors of Yakutsk array.}
\label{PetrovIS-fig1}
\end{figure}

In the mid 80-es of the last century, the Radio Array with registration bandwidth 30-40 MHz was designed as an extension of main Yakutsk particle array [5]. Air shower radio emission is registered by 20 receiving antennas, which are installed on 10 pillars as shown in Fig. ~\ref{PetrovIS-fig1}.

One pillar consists two independent half-wave dipoles with orientation E-W and N-S. Antennas installed at $\lambda$ / 4 above ground, thus ensuring a maximum of the radiation pattern for the emission coming from the top.

In the early 2009 at Yakutsk array the radio array was re-created [10]. The array consisted 12 crossed at 90 $^\circ$ receiving antennas, oriented in E-W and  N-S, peripheral recording device and PC storage data. The recording device was located directly underneath of antenna. Antennas are deployed near center of Yakutsk array and consist 2 independent clusters that synchronized by GPS. The distances from the center are 50-100 m.

\section{Results}

Over 10 years of operation of the Radio array, observation time is 50400 hours. 14700 events of air showers have been reported with the radio emission, whose axis is in a circle with a radius of 0.5 km and energy exceeding 10$^{17}$ eV. During this time, showers with energy exceeding 10$^{19}$ eV were registered. From this data set we found a correlation with air showers parameters [9] and demonstrated the possibility of using EAS radio emission to study the physics and astrophysics of ultra-high energy cosmic rays [11].

\begin{figure}[h]
\begin{minipage}[h]{0.8\linewidth}
\center{\includegraphics[width=0.78\linewidth]{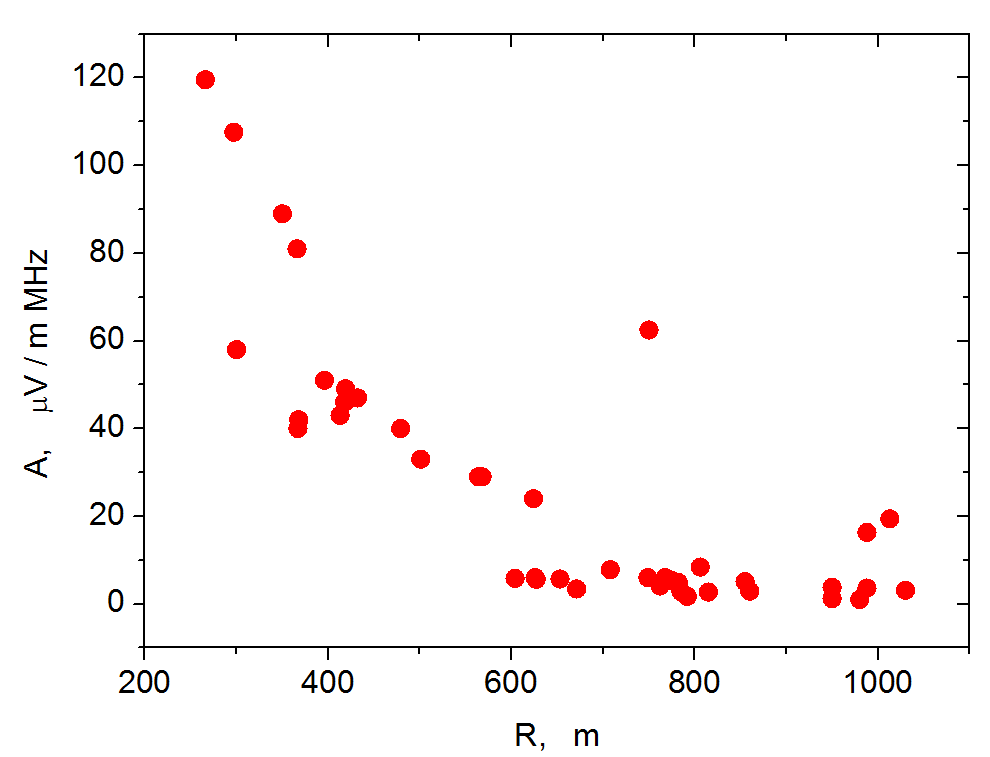} \\ a)}
\end{minipage}
\hfill
\begin{minipage}[h]{0.8\linewidth}
\center{\includegraphics[width=0.78\linewidth]{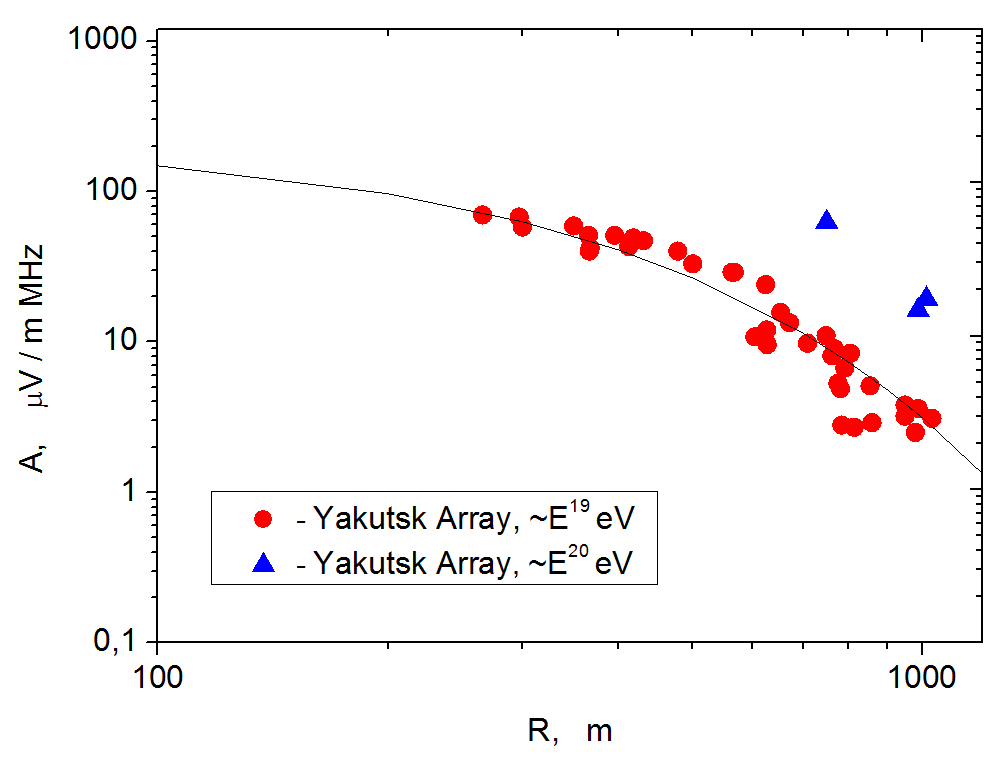} \\ b)}
\end{minipage}
\caption{ Lateral distribution of electromagnetic filed strength  from the distance to air showers axis. a) in metric representation b) in logarithmic representation}
\label{PetrovIS-fig2}
\end{figure}

Fig. ~\ref{PetrovIS-fig2} shows the cloud of points. Most of the showers has energies from 10$^{19}$ to 4$\cdot$10$^{19}$ eV. Noteworthy that some points located considerably higher than the total weight of the points. We have not normalized the data for energy purposely in order to emphasize the special status of these points, namely, they are two showers with energy $\geq$ 10$^{20}$ eV and thus have almost an order of magnitude greater amplitude than the rest of the showers.

As can be seen from Fig. 2a, obtained dependence of the amplitude of a radio signal from a distance is close to the simple exponential mean of the type (~\ref{IP_eq1}):

\begin{equation}\label{IP_eq1}
E = \varepsilon\cdot\exp\left(\frac{R}{R_0}\right)
\end{equation}

In Fig. 2b points are normalized to average energy $<$E$_0$$>$ = 3.7$\cdot$10$^{19}$ eV and average zenith angle $<\theta>$ = 43.1$^\circ$. It is seen that after the normalization the cloud of points indicates the rapid decay of radio signal and reflects the function of spatial distribution of radio signal depending on the distance from an antenna to air shower axis.

We derived formula to fit experimental data (~\ref{IP_eq2}), which in Fig. 2b. is shown as solid curve. The fit explains well data on medium and large distances from shower axis.

\begin{eqnarray}\label{IP_eq2}
  \varepsilon = (29.5\pm1.6)\left(\frac{E}{5\cdot10^{17}}\right)^{0.83\pm0.03}
  \nonumber\\
  \times(1-\cos\theta)^{1.16\pm0.03} \\
  \times\left\{
   - \frac{R}{162\pm8 + (84\pm3)
   \left[
   \frac
   {X-675}
   {100}
   \right]}
    \right\}  \nonumber
\end{eqnarray}

\begin{equation}\label{IP_eq3}
  <\ln A> =
  \frac
  {P^{exp} - P^p}
  {P^{Fe} - P^p}
  \cdot \ln A_{Fe}
\end{equation}

Here P$_{i}$ – parameter that characterizes longitudinal development of air shower X$_{max}$.

Using the interpolation method (~\ref{IP_eq3}) [12], and model calculations QGSJETII-04 for the proton and iron core evaluation of the atomic weight of primary cosmic ray particles that produced EAS with such energy were made. The value of $<lnA>$ = 1.5$\pm$0,7, which corresponds to radiation consisting of helium nuclei and average type CNO nuclei [13]. Therefore registration of radio emission of the giant energy air showers can be independent method to learn physics as well as astrophysics of cosmic radiation.

\section{Conclusion}

Long-term observation of radio emission at the Yakutsk array proved the existence of radio emission at energies 10$^{19}$ eV and it allowed obtaining the characteristics of radio emission at such energies:
\begin{itemize}
\item[a] Attenuation function of air shower radio signal from distance at energies 3.7$\cdot$10$^{19}$ eV and its gradient;
\item[b] Proved the existence of radio emission at energies 10$^{20}$ eV in most energetic air showers that were registered at Yakutsk array [14];
\item[c] Significant signal in inclined showers and influence of magnetic field on LDF shape, which confirms once more [9] a role of geomagnetic mechanism of radio emission generation;
\item[d] Estimation of mass composition of cosmic ray obtained in the frame of QGSJETII-04 model by radio data at energies above 10$^{19}$ eV in a good agreement with our results presented in [12, 13] and points on composition mostly He and CNO nuclei. This result does not contradict the result of the mass composition obtained at huge EAS arrays [15].

\end{itemize}

\bigskip 
\begin{acknowledgments}
The research supported by the Russian Foundation for Basic Research (grant 16-29-13019 ofi$\_$m).
\end{acknowledgments}

\bigskip 

\end{document}